\documentclass[12pt]{article}
\usepackage{amssymb,amsfonts}
\newcommand{\be}{\begin{equation}}
\newcommand{\ee}{\end{equation}}
\newcommand{\bea}{\begin{eqnarray}}
\newcommand{\eea}{\end{eqnarray}}
\newcommand{\bref}[1]{(\ref{#1})}

\def\ie{{\it i.e. }}

\def\t{\theta}

\def\del3{\delta^{(3)}}

\def\undos{{1 \over 2}}

\newcommand{\C}[1]{{\cal #1}}

\def\e{\epsilon}
\def\otaula{\begin{tabular}}
\def\ctaula{\end{tabular}}
\def\espai{\;\;\;\;\;\;}
\def\zespai{\;\;\;\;}

\def\G{\Gamma}

\def\gym{g_{YM}}

\def\a{\alpha}
\def\b{\beta}

\def\l{\lambda}
\def\g{\gamma}

\def\te{\tilde{\epsilon}}
\def\tte{\tilde{\tilde{\epsilon}}}

\def\te{\tilde{\epsilon}}

\def\avall{\vspace{1cm}}
\def\zavall{\vspace{0.4cm}}
\def\ka{K\"ahler }
\def\et{\tilde{e}}
\def\dtau{\dot{\tau}}
\def\dsigma{\dot{\sigma}}

\def\T{\Theta}
\def\CN{$\mathcal{N}$}
\def\tH{\tilde{H}}
\def\th{\tilde{h}}

\def\we{\wedge}
\def\MR{\mathbb{R}}
\def\MC{\mathbb{C}}

\begin{document}

\begin{titlepage}

\setcounter{page}{0}
\begin{flushright}
ECM-UB-02/14 \\
DFF-391-07-02\\
\end{flushright}

\vspace{5mm}
\begin{center}
{\Large {\bf Supergravity Duals of Noncommutative Wrapped $D6$ Branes and
Supersymmetry without Supersymmetry}}
\vspace{10mm}

{\large Jan Brugu\'es$^a$, Joaquim Gomis$^{a,b}$, Toni Mateos$^a$, 
Toni Ram\'{\i}rez}$^a$ \\
\vspace{5mm}
${}^a${\em Departament ECM, Facultat de F{\'\i}sica, \\
Universitat de Barcelona and Institut de F{\'\i}sica d'Altes Energies, \\
Diagonal 647, E-08028 Barcelona, Spain }\\
\vspace{5mm}
${}^b${\em Dipartimento di Fisica, Universit\`a di Firenze \\
INFN, Sezione di Firenze, I-50125 Firenze} 
\end{center}
\vspace{20mm}

\centerline{{\bf{Abstract}}}

\vspace{5mm}

We construct the supergravity solution in 11 dimensions describing D6-branes
wrapped around a \ka four-cycle with a $B$-field along the flat
directions of the brane. The configuration is dual to an ${\cal N}$=2
noncommutative gauge theory in $2+1$ dimensions.
We also construct the four associated
independent Killing
spinors. The phenomenon of supersymmetry without supersymmetry
appears naturally when compactifying to type IIA or 8d gauged
supergravity. Therefore, this solution also provides an 11d background
with four supercharges and four-form flux, which is not
obtainable from 8d gauged supergravity.

\vfill{
 \hrule width 5.cm
\vskip 2.mm
{\small
\noindent E-mail: jan,gomis,tonim,tonir@ecm.ub.es \\
}}

\end{titlepage}
\newpage
\section{Introduction}

In the last two years, an extension of the AdS/CFT \cite{Maldacena:1997re}
\cite{Gubser:1998bc}
\cite{Witten:1998qj}
 correspondence
to gauge field theories with less than maximal supersymmetry has been
achieved via wrapped branes,
and many supergravity duals have been constructed,
see $e.g.$ \cite{Maldacena:2000mw}
-\cite{Naka:2002jz}.
Gauged supergravities provided a useful tool to construct such configurations
since they geometrically implement the twisting  
\cite{Bershadsky:1995qy} of the field theories
in a natural way.
These supergravity solutions have been used to study qualitative aspects of
non-perturbative gauge theories like quark confinement, chiral
symmetry breaking, supersymmetry breaking and renormalization group flow.

On the other hand,
supergravity duals of noncommutative (NC) theories with
maximal supersymmetry have also been constructed 
\cite{Hashimoto:1999ut}-
\cite{Berman:2000jw}
by switching on a background $B$-field. See 
\cite{Larsson:2001wt}
for a 
summary of all Dp-brane solutions.

The aim of this paper is to join both ideas and produce
supergravity duals of non-maximally supersymmetric NC
field theories by obtaining solutions of wrapped branes in
a nontrivial $B$-field background. In particular, we will
construct the sugra dual of an \CN=2 NC field theory in 2+1
dimensions by wrapping D6-branes around a \ka four-cycle
\footnote{We recall that in the presence of $B$-fields,
the unwrapped D6-branes are expected to be free
from the usual bulk gravity decoupling problems.\cite{Jabbari}}.

As we show, such solutions could not have been obtained using the
corresponding 8d gauged supergravity 
\cite{Salam:1984ft}, since even the
unwrapped NC configuration looses all supersymmetry in the
compactification. This is the well-known phenomenon of
supersymmetry without supersymmetry \cite{Duff:1997qz}, and it forces
us to work directly in 11d sugra \cite{Cremmer:1978km}.

In order to illustrate clearly this phenomenon, we first reconsider
the unwrapped NC D6 in 11d which, unlike the
commutative case, involves a non-trivial four-form flux. We give an ansatz,
we derive and solve the first order BPS equations and, most importantly,
we obtain the explicit form of the 16 Killing spinors. These allow us
to interpret the solution as a non-threshold bound state of MKK-M5 (or
D6-D4 in IIA), and to show which of its possible compactifications
to type IIA or 8d gauged sugra are supersymmetric.

We then apply the same method to the case of N D6 wrapping a
calibrated \ka four-cycle inside a Calabi-Yau three-fold. The 11d
description (obtained in \cite{Gomis:2001vg}) was purely gravitational
and it was given by a 3d Minkowski spacetime times a Calabi-Yau four-fold.
The metric was constructed in 8d sugra by demanding supersymmetry and
then uplifted to 11d.
We will construct the NC version of such configuration. The Killing
spinors will allow us to interpret it as a non-threshold bound state of
MKK-M5 (or D4-D6 in type IIA) with the D4 completely wrapped around
the \ka four-cycle. This background has 4 linearly realised supersymmetries
and is dual (in the IR) to an \CN=2 $U$(N) NC field theory
in 2+1 dimensions. Since the noncommutativity is along the
two spatial directions, such theory is nonlocal in space, but
free from unitarity or causality problems \cite{Gomismehen, Seiberg}.

The organization of the paper is as follows. In section 2 we
reconsider the case of the unwrapped NC D6-branes and
illustrate how supersymmetry without supersymmetry arises in compactifying.
In section 3 we consider the case of the NC wrapped D6-branes, we
obtain the new 11d supergravity solution and discuss its properties
and its supersymmetry.
In section 4 we give some conclusions.

\section{Flat noncommutative D6-branes}

\subsection{Obtaining the solution in 11d}

The purpose of this section is to find the supergravity solution
describing N flat D6-branes in the background of a magnetic $B$-field.
We will call this configuration a NC flat D6 brane.

The M-theory description \cite{Townsend:1995kk,Witten:1995ex}
of the commutative D6 is simply given
by pure geometry, 
so we will work in 11d Sugra.
Nevertheless, since
we will turn on a $B$-field in the IIA description, we will need to
turn on a three-form  to describe the NC case. We will first obtain
the whole geometry and then take the near horizon limit.

The first step will be to make and ansatz for the bosonic fields (metric
and $A_{[3]}$) of 11d sugra. In order to do so, recall that the 11d solution
for N flat commutative D6-branes is the product of $\MR^6$ with the Euclidean
Taub-Nut space \cite{Gibbons:zt}
\be
\label{flatcase} ds^2_{(11)}= dx^2_{0,6}+H \left(dr^2+{r^2}
[d\t^2+sin^2 \t d\phi^2]\right)+ R^2 H^{-1} \left( d\psi+cos\t d\phi
\right)^2, 
\ee
 with
\be
H(r)=1+{R\over r}, \espai\espai R=g_s N \sqrt{\a'},
\ee
and N is the number of D6-branes.
\footnote{The coordinates range is $0 \leq \t < \pi$, $0 \leq \phi < 2\pi$
and $0 \leq \psi < {4\pi\over N}$, the latter demanded by regularity of
the metric about the origin.}

 The Taub-Nut space is a $U$(1) bundle
over $\MR^3$ and it turns out to be a hyper-\ka manifold with $SU$(2) holonomy.
It is remarkable that the near horizon limit of this solution
gives the product of $\MR^6$ with an $A_{N-1}$ ALE singularity,
which is nothing but the orbifold $\MC ^2/Z_N$. Apart from the
global identifications imposed when modding by $Z_N$, the metric
is locally flat. Therefore, any analysis based only on local
properties of the manifold will not be able to distinguish the
actual manifold from flat space.

Now we would like to make a noncommutative deformation by turning
on a $B$-field (in a IIA description) along the $(x^5,x^6)$ plane.
This will explicitely break the  $SO(1,6)$ isometry of the
worldvolume to an $SO(1,4)\times SO(2)$, so that the ansatz for
the metric is 
$$ ds^2_{(11)}=\tau^2(r)\left[
dx^2_{0,4}+ \sigma^2(r) dx^2_{5,6}+H\left(dr^2+ r^2
[d\t^2+sin^2 \t d\phi^2] \right)\right]+$$
\be\label{ans}
+ \tau^{-4}(r) R^2 H^{-1}
\left(d\psi+cos\t d\phi \right)^2. \ee

Note that the factor in front of the $U$(1) fiber is related
to the one in front of the other ten coordinates through the 
ansatz for lifting IIA solutions to M-theory \cite{Campbell,Huq}:
\be \label{lift1}
ds^2_{(11)}=e^{-2\Phi/3}ds^2_{IIA}+e^{4\Phi/3}\left(dx^T+C_{[1]}\right)^2,
\ee
where $x^T$ is the M-theory coordinate.
We also make an ansatz for
the three-form that respects the $U$(1) monopole fibration 
\be
A_{[3]}=\chi(r)\,\, dx^5 \we dx^6 \we (d\psi+cos\t d\phi). \ee 
We will determine the functions of our ansatz by demanding that the
supersymmetry transformations admit a non-trivial Killing spinor.
Since the background is bosonic, we just need to care about the
gravitino variation
\begin{equation} \label{gravitino}
\delta \Psi _A  =  D_A\epsilon -\frac{1}{288}
\left( \Gamma _A{}^{BCDE}-8\delta _A{}^{[B}\Gamma ^{CDE]}\right)F_{BCDE}
 \epsilon,
\end{equation}
where $D_A= (\partial_A+ \frac 14 \omega_A{}^{CD}\Gamma_{CD})$ 
is the covariant derivative in flat coordinates and $F_{BCDE}$ is the four 
field strength form.

In what follows it will be very important to make clear the
vielbein basis that we are using, since the explicit form of the
Killing spinors depends on it. We choose the following vielbein
for the diagonal part of \bref{ans}
\be
e^a=\tau(r)\,dx^a\, , \,\, a=0,..,4 \espai\espai
e^{i}=\tau(r)\sigma(r)\, dx^{i}\, , \,\, i=5,6 \espai\espai
 e^7=\tau(r)H^{\undos}(r) \, dr,
\ee
while for the squashed $S^3$ we take
\be
\label{vi} e^8=\tau(r) H^{\undos}(r) r \, \et^1 \espai\espai
e^9=\tau(r) H^{\undos}(r) r \, \et^2 \espai\espai e^T=
\tau^{-2}(r)H^{-\undos}(r)R
\, \et^3, \ee with $\et^i$ the typical vielbeins of a round $S^3$ \be
\et^1=d \t \espai\espai \et^2=sin\t d\phi \espai\espai
\et^3=d\psi+cos\t d\phi. \ee 
Now we proceed to analyse the
supersymmetry variations. Due to the $SO(1,4)$ symmetry, the
equations for $A=0,1,2,3,4$ are equivalent. If we assume  that
the Killing spinors do not depend on the coordinates
$\{x^0,...,x^6\}$, these equations can be written as
\be\label{gamma1}
 \left( cos\alpha \G_{D6}+sin\alpha \G_{D4}\right) \e
=- \e, \ee
with
\be\label{cos}
 cos\alpha \,= \, {\chi'\over \chi} \tau^3 
H R^{-1}r^2 \zespai\zespai sin\alpha \,= \, -6 \tau^3 \chi^{-1}
\tau' \sigma^2 H^{\undos} r^2 \zespai\zespai \Gamma_{D6}\equiv \Gamma_{0123456}
\zespai\zespai \Gamma_{D4}\equiv\Gamma_{01234T}.
\ee
Since $\{\G_{D6},\G_{D4}\}=0$,
equation \bref{gamma1} is telling us that we are obtaining a
non-threshold bound state of D6-D4 from a IIA point of view, or
a bound state MKK-M5 from an M-theory one \cite{Townsendalg}.
To proceed, note that the equation \bref{gamma1}can be rewritten as
\footnote{For a different configuration of M branes an analogous technique was
used in  \cite{Camino:2001jg}.}

\be\label{gamma2}
 e^{-\alpha \G_{56T}} \e=-\G_{D6}\e, \ee
whose most general solution is
\be \label{analog}
 \e\,\,=\,e^{{\alpha \over 2}\G_{56T}}\, \te(r,\t,\phi,\psi),
\espai\espai
with \espai \espai
\G_{D6}\te(r,\t,\phi,\psi)\,=\,-\te(r,\t,\phi,\psi).
\ee
Note that
the angle $\alpha$ is a function of $r$. At this point we need to make
an ansatz for $\te$,

\be
\label{spinorans}
\te(r,\t,\phi)=
f(r)e^{{\t \over 2}\G_{78}}
e^{{\phi \over 2}\G_{89}} \e_0,
\ee
where $\e_0$ is a constant spinor verifying $\G_{D6}\,\e_0\,=\,-\e_0$.

One can plug our ansatz in the remaining supersymmetry variations and obtain
the following set of first order, coupled, non-linear BPS equations

\be
3{\tau' \over \tau}+{\sigma'\over \sigma} =  0
\ee\be
\chi\chi'-6R^2H^{-1}\sigma^4 {\tau'\over \tau} = 0 
\ee\be
{3\tau'\over \tau} +{\chi'\over 2\chi}+{H'\over 2H} = 0 \label{tre}.
\ee
The general solution depends
on a total of three arbitrary constants. Two of them can be
fixed by demanding that the solution reduces to the commutative
one \bref{flatcase} when the $A_{[3]}$ is set to zero
(commutative limit). 
The remaining arbitrary constant has a
physical meaning: it is the strength of the noncommutativity,
that we call $\T$. We have
\be
\tau(r)=h^{{1\over 6}}, \espai\espai \sigma(r)=h^{-\undos},
\espai\espai 
\chi(r)=-{\T R \over H h}, \espai\espai f(r)=h^{1\over 12}(r),
\ee
 where we have defined
\be
h(r)=1+\T^2 H^{-1}.
\ee
Summarising, the 11d metric, 3-form and the Killing spinors are given by
\zavall
\be\label{nonnearh}
 ds^2_{(11)}= h^{1\over 3}\left(-dx_{0,4}^2+
h^{-1} dx_{5,6}^2
+ H [dr^2+r^2 d\Omega_2^2] \right)+ 
H^{-1}h^{-2/3}R \left( d\psi +cos\t d\phi \right)^2 \ee\be
A_{[3]}=-{\T R \over Hh} \;\; dx^5 \wedge dx^6\wedge \left(
d\psi+cos\t d\phi\right) \espai\espai \label{spinsol1}
\e(r,\t,\phi,\psi)=h^{1\over 12}(r) e^{{\alpha (r)\over 2}
\G_{56T}} e^{{\theta\over 2} \G_{78}} e^{{\phi\over 2}\G_{89}}
\e_0 \ee with \be \G_{D6}\e_0=-{\e}_0 \espai \espai cos\alpha =
h^{-1/2} \espai\espai sin\alpha=\T (Hh)^{-\undos}. \ee

\zavall

This solution describes the whole geometry of N flat NC D6-branes
and the number of independent Killing spinors is 16. The
configuration  corresponds to a bound state of N MKK monopoles
and N M5 branes, or a bound state of N D6-D4 branes in type IIA.
If we want to use this background, in the spirit of the AdS/CFT
\cite{Maldacena:1997re}\cite{Gubser:1998bc}\cite{Witten:1998qj}
correspondence, to study the dual NC field theory, we must take
the near horizon limit, which consists of taking $\alpha'
\rightarrow 0$ keeping fixed 
\be \label{nhlimit} u={r\over
\alpha'}, \espai \espai \tilde{\T}=\alpha' \T, \espai \espai
g^2_{YM}= g (\alpha')^{3/2}. \ee 
After a suitable change of
variables, \footnote{Explicitely $u={y^2 \over 4 N \gym}$.} the
metric and the three-form become \cite{Larsson:2001wt}

\be \label{nearh} ds^2_{11}= h^{1/3} \left[dx_{0,4}^2 +
 h^{-{1}} dx_{5,6}^2 +
dy^2+ {y^2\over 4} \left( d\Omega_{(2)}^2 +
h^{-1} [ d\psi+cos\theta d\phi ]^2 \right)\right] \ee \be
A_{[3]}= -{\tilde{\T} \over 4N\gym^2} \, {y^2\over h} \,\, dx^5\wedge
dx^6 \wedge \left( d\psi +cos\theta d\phi\right) \espai\espai
h(y)=1+\left({ \tilde{\T}\,\, y \over 2N\gym^2}\right)^2. \ee The number of
supersymmetries continues to be 16 so, unlike the commutative
case, there is no enhancement. Note that this
background should be in the IR
the large N dual of a 6+1 noncommutative gauge theory with 16
supercharges.

Finally, we would like to consider the commutative limit of
\bref{nonnearh} and \bref{nearh}. Sending $\T \rightarrow 0$
implies $h \rightarrow 1$. In such limit, eq.\bref{nonnearh}
collapses to eq.\bref{flatcase} and the 16 spinors become simply
\be \label{spinsol2} \e(\t,\phi)= e^{{\theta\over 2} \G_{78}}
e^{{\phi\over 2}\G_{89}} \e_0, \espai\espai with \espai\espai
\G_{D6}\e_0=-{\e}_0. \ee On the other hand, in the commutative
limit, eq.\bref{nearh} becomes the aforementioned $A_{N-1}$
singularity. Apart from the previous 16 spinors, it also admits
the following 16 ones \be \label{spinsol3}
\e(\psi)=e^{-{\psi\over 2}\G_{89}}\e_0, \espai\espai with
\espai\espai \G_{D6}\e_0=\e_0 \label{enhance}. \ee Note that they
have a different eigenvalue with respect to $\G_{D6}$. Modding
out by the $Z_N$ global identifications brings the number of
supersymmetries back to 16. Only for N=1, flat space, we
have a true enhancement of susy.

\subsection{Compactification to type IIA and Supersymmetry}
\label{susywsusy}

As is well known, any configuration of 11d supergravity can be
consistently reduced to 10d IIA supergravity as long as it is a
$U$(1) fibration over a ten-dimensional base space \cite{Nilsson}.
The bosonic part of the reduction ansatz is \cite{Campbell,Huq}
\be \label{reduction} ds^2_{(11)}= e^{-{2\phi\over 3}} ds^2_{IIA}
+ e^{4\phi\over 3}\left( dx^{T}+C_{[1]}\right)^2, \ee \be
A_{[3]}=- C_{[3]} + dx^{T} \wedge B_{[2]}, \ee 
where $\partial_{x^T}$ is the Killing vector that generates
the $U$(1) isometry.

 Indeed, the
ansatz does not finish here. First of all, the vielbeins must be
selected so that the first ten of them do not depend on $x^{T}$
\footnote{This statement can be made more rigorous by computing
the more intrinsic Lie-Lorentz derivative \cite{Ortin:2002qb} with
respect to the Killing vectors. For the cases considered in this paper,
such derivative collapses to the usual one.}.
Second, once we fix the vielbein, the supersymmetry parameter
must also be independent of $x^{T}$. These comments are
relevant, since we will show that some of our configurations fit
perfectly into the ansatz \bref{reduction} but do not verify the
condition on the susy parameter. In such cases, one produces
solutions of the $IIA$ equations of motion, but they do not
typically have any linearly realised supersymmetry 
\cite{Bergshoeff:1994cb}. Note as well,
that all the metrics in this section have at least two different
$U$(1) isometries, generated by the Killing vectors
$\partial_{\psi}$ and $\partial_{\phi}$. The amount of
supersymmetry preserved in the reduction to $IIA$-sugra, will
depend on whether we have noncommutativity or not.

\avall

{\bf I. Noncommutative case:} We will work with the basis \bref{vi}.
Since it does not explicitely depend on $\phi$ nor $\psi$, it can
be used for the reduction along both $U$(1) isometries.
Fixed the vielbeins,
we see that, in both the whole
geometry \bref{nonnearh} and in the near horizon \bref{nearh},
the sixteen spinors depend on $\phi$ but not on $\psi$.
As a consequence, reducing along the $\partial_{\psi}$ isometry will
preserve the whole 16 supercharges, and we obtain the same
supersymmetric configuration found in {\it e.g.} \cite{Larsson:2001wt}.

If we want to reduce along $\partial_{\phi}$, we will produce a solution
of type IIA supergravity which will not be supersymmetric. In order to do so,
we need to reexpress our metric \bref{nonnearh} in a form that makes
the new $U$(1) fibration more explicit
\be
ds^2_{(11)}=h^{1\over 3}\left(dx_{0,4}^2+h^{-1}dx_{5,6}^2+ H[dr^2+r^2
d\t^2]\right)+
h^{-{1\over 3}}\l^{-1}R r^2 sin^2\t \, d\psi^2
+\l\left(d\phi+A_{[1]}\right)^2
\ee
with
\be A_{[1]}\equiv \l^{-1}H^{-1}h^{-{2\over 3}}Rcos\t \, d\psi, \espai\espai
\l(r,\t)\equiv Hh^{1\over 3}r^2 sin^2\t+H^{-1}h^{-{2\over
3}}Rcos^2\t.
\ee
So reducing along $\partial_{\phi}$ gives
\be
\label{pordios} 
ds^2_{IIA}=\l^{\undos}
h^{1\over 3}\left(dx_{0,4}^2+h^{-1}dx_{5,6}^2+ H[dr^2+r^2
d\t^2]\right)
+\l^{-\undos}h^{-{1\over 3}}R r^2 sin^2\t \, d\psi^2
\ee\be
e^{4\Phi/3}= \l(r,\t) \espai\espai B_{[2]}=-{\Theta R\over Hh}cos\t \,
dx^5\wedge dx^6 \ee
\be C_{[1]}=A_{[1]} \espai\espai
C_{[3]}={\Theta R \over Hh}\, dx^5\wedge dx^6 \wedge d\psi.
\ee
It can be checked that such configuration is a solution of the 
type IIA equations of motion and that it does not preserve any
supercharge.
This is an example of the phenomenon of supersymmetry without
supersymmetry \cite{Duff:1997qz}.

\avall

{\bf II. Commutative case:} We can obtain the commutative configurations
just by letting $\Theta \rightarrow 0$ in the former equations. 
When reducing along $\psi$, this just gives the usual 
D6-brane metrics of type IIA. Nevertheless, when reducing along $\phi$ we have to
distinguish whether we are or not in the near horizon limit.
If we are still in the whole geometry, the only Killing spinors are
those in \bref{spinsol2}, which explicitely depend on $\phi$. So the
$\partial_{\phi}$-reduction kills all supersymmetry, and it
simply gives \bref{pordios} with the replacement $h\rightarrow 1$.

The major change comes in the near horizon limit of the commutative case
which, recall, it is just locally flat space in 11d. As we said, there 
is a local enhancement to 32 supersymmetries, and sixteen new spinors
appear. They are the ones of \bref{enhance}, and they depend only on $\psi$.
Therefore, they all survive the $\partial_{\phi}$-reduction and must 
produce locally a 16-supersymmetric configuration.
By construction, this solution is nothing but the near horizon limit
of \bref{pordios} with $h=1$. The expected 16 Killing spinors
can simply be obtained using the reduction formulas \cite{Campbell,Huq}
\be
\e(\psi,\t)=e^{\Phi \over 6} e^{-{\psi\over 2}\G_{89}}\e_0,
\espai\espai with \espai\espai \G_{D6}\e_0=\e_0. \ee

\subsection{Compactification to 8d gauged sugra and Supersymmetry}
\label{susywsusy2}

Gauged supergravities have recently been exploited to obtain
configurations of wrapped branes, since they provide a natural
method to implement geometrically the twistings of the 
supersymmetric field
theories\cite{Bershadsky:1995qy}. In this subsection we show that one runs
into trouble when trying to use them to obtain NC duals via
wrapped branes. In particular, we will show that the NC flat
configurations, both for the whole geometry \bref{nonnearh} and
for the near-horizon limit \bref{nearh}, are not supersymmetric
from the point of view of 8d gauged sugra.
The compactification of
M-theory on an $SU$(2) manifold was worked out in \cite{Salam:1984ft},
and it gave a maximal 8d $SU$(2) gauged supergravity. This theory
was used in 
\cite{Edelstein:2001pu,Hernandez:2001bh,Gomis:2001vg,Gursoy:2002tx,Hernandez:2002ig}
to obtain sugra duals of non-maximally supersymmetric field
theories. Now we would like to see if our configuration
\bref{nonnearh} and \bref{nearh} could have been found by using
this 8d gauged sugra. To answer this question, the first thing
to do is to choose the vielbein that is implicit in \cite{Salam:1984ft}.
In particular one needs to work with the $SU$(2) left invariant
one-forms for the squashed $S^3$ part of the metric. So that,
instead of \bref{vi}, one should use \be \label{vi2}
\hat{e}^8=\tau(r) H^{\undos}(r) r \, w^1 \espai\espai
\hat{e}^9=\tau(r) H^{\undos}(r) r \, w^2 \espai\espai
\hat{e}^T=\tau^{-2}(r)H^{-\undos}(r)R\, w^3, \ee 
with \footnote{Note that the signs
have been chosen so that both basis share the same orientation.}
\be w_1=-cos\psi d\theta - sin\theta sin\psi d\phi \zespai\zespai
w_2=-sin\psi d\theta + sin\theta cos\psi d\phi \zespai\zespai
w_3=-d\psi-cos\theta d\phi. \label{vi3} \ee 
We will call \bref{vi2} the
$w$-base and \bref{vi} the $e$-base. It is easy to work out the
form of the spinor in this new base, since we have just performed
a local lorentz transformation. Explicitely, it can be seen that
the $w$-base can be obtained from the $e$-base by performing a
rotation of $\pi$ along $x^9$, followed by a rotation of angle
$-\psi$ along $x^T$. So the Killing spinors will transform with
the (inverse) spin $\undos$ representation of such rotations \be
\label{rotation} \e'\,=\,e^{-\psi {\G_{89}\over
2}}e^{\pi{\G_{T8}\over 2}}\,\e\,=\, \G_{T8}e^{\psi {\G_{89}\over 2}}\,\e.
\ee

We can now see that all the Killing spinors obtained in 11d,
after the change of base, become $\{\t,\phi,\psi\}$-dependent,
while only the 16 ones that produced the enhancement in the
commutative near-horizon limit \bref{enhance} become constant
spinors.

The compactification on a group manifold \cite{Scherk}
assumes that supersymmetry parameters do not depend on the internal space coordinates.
So we can already predict
that only the commutative near-horizon limit will appear to be supersymmetric
in 8d. Indeed, this made possible the obtention of a stack of N 
parallel flat D6-branes
by analysing the BPS equations of 8d \cite{Townsenddom}.
On the contrary, the NC configurations could never have been
obtained by making an ansatz in 8d and looking at the susy variations.
As an example, the configuration \bref{nearh} can be reduced to 8d since it
fits into the bosonic ansatz \cite{Salam:1984ft}, and produces

\be \label{one}
ds^2_{(8)}={g\over 4}y \,h^{1/3} \left( dx_{0,4}^2+h^{-1} dx_{5,6}^2
+ dy^2\right)
\ee
\be
e^{2\phi\over 3}={g\over4} y \espai\espai e^{\lambda}=h^{1/6}
\ee
\be \label{two}
G_{[2]}=-{\T g^2\over 16 N \gym^2}\,{y^2 \over h} \,\, dx^5\wedge dx^6
\espai\espai G_{[3]}=-{\T g \over 4 N\gym^2}{y\over h^2} \,\,
 dx^5\wedge dx^6\wedge dy,
\ee
where $\lambda$ is a scalar field  on the coset space $\frac{SL(3,R)}{SO(3)}$
and $G_{[2]}$ and $G_{[3]}$ are field strength forms of 8d Sugra.
This is again a solution of the 8d equations of motion but,
as can be explicitely seen, it is not supersymmetric.
In the next section we will obtain the wrapped version of all
these configurations, and we will apply the same arguments
to proof that the NC cases could not have been found from 8d sugra.

\section{Noncommutative Wrapped D6-branes}

The configuration of N D6-branes wrapping a \ka four-cycle inside a
Calabi-Yau three-fold was obtained in \cite{Gomis:2001vg}. In this
section we will first discuss some issues of its supersymmetry properties
and then we will obtain its NC deformation.

\subsection {Commutative wrapped D6-branes}

By using 8d gauged supergravity, the purely
gravitational 11d description of such configurations was obtained in
\cite{Gomis:2001vg}
\be\label{kahler1}
ds^2_{(11)}=dx_{0,2}^2+{3\over
2}(r^2+l^2)ds^2_{cycle}+U^{-1}dr^2+ {r^2 \over 4}\left( d\t^2+
sin^2\t d\phi^2\right)+{1\over 4}U r^2\sigma^2, \ee
where $\sigma$ described the following $U$(1) fibration
\be \sigma= d\psi+cos\t d\phi + A_{[1]}
\espai\espai
with \espai\espai
dA_{[1]}=6 J_{[2]}. \ee
Here, $J_{[2]}$ is the \ka form of the four-cycle chosen and
$U(r)={ 3r^4+8l^2r^2+6l^4 \over 6(r^2+l^2)^2}$.
This solution has the topology of
$\MR^{1,2}\times CY_4$, the Calabi-Yau four-fold being a $\MC^2$
bundle over the \ka four-cycle $\C{K}_4$ \footnote{The metric for
this Calabi-Yau four-fold was first found in \cite{gibbonsetal} in
a completely different approach.}, \footnote{This construction exemplifies
the uplift from a manifold with $SU(3)$ holonomy in type IIA to a manifold 
with $SU(4)$ holonomy in M Theory 
\cite{Gomis:2001vk}.}.
 Alternatively, it can be seen
as a cone with $r=cte$ hypersurfaces described by a $U$(1) bundle
over the base $S^2\times \C{K}_4$.
\footnote{For simplicity, in this paper we will only consider
the four-cycle $\C{K}_4=S^2\times S^2$, although the results can be
generalised to any other choice. So, in this case,
$ds^2_{cycle}={1\over 6}\left(d\t_1^2+sin^2\t_1 \, d\phi_1^2+
d\t_2^2+sin^2\t_2 \, d\phi_2^2\right)$ and $A_{[1]}=cos\t_1 d\phi_1+cos\t_2 d\phi_2$.}
In the following we choose the vielbeins so that the \ka form of the such base
is written as $J_{[2]}=e^3\wedge e^6+e^4\wedge e^5+e^8\wedge e^9$.

Performing the same analysis of the supersymmetry variations
again, one finds the following Killing spinors
\be \label{mmm}
\e(\psi)=e^{-{\psi\over 2}\G_{89}}\e_0,
\ee
with $\e_0$ a constant spinor subject to
\be
\G_{D6} \,\e_0\,=\,\e_0, \espai \espai \G_{36}\,\e_0\,=\,\G_{45}\,\e_0\,=\,\G_{89}\,\e_0.
\ee
The first condition just signals the presence of the D6, while the other two
are the usual projections of a Calabi-Yau three-fold. 
Altogether, the configuration preserves only ${1/8}$ of the 32 supersymmetries.

In \cite{Gomis:2001vg} the compactification to IIA was performed along
the immediate $U$(1) isometry generated by ${\partial_{\psi}}$.
This lead to the following type IIA configuration \footnote{There is a 
typo in eq (28) of  \cite{Gomis:2001vg}.}
\be \label{mal}
ds^2_{IIA}=e^{2\Phi/3}\left[dx_{0,2}^2+{3\over
2}(r^2+l^2)ds^2_{cycle}+U^{-1}dr^2+ {r^2 \over 4}\left( d\t^2+
sin^2\t d\phi^2\right)\right]\ee
\be e^{4\Phi/3}=U(r)r^2 \espai\espai
C_{[1]}= \undos \left(cos\t d\phi + A_{[1]}\right).\ee
A probe brane analysis showed that there was no moduli space.
This can now be understood, since the Killing spinors \bref{mmm}
depend on the compactification coordinate and, therefore, do not
fit in the reduction ansatz. Thus, as can be checked explicitely,
\bref{mal} is a solution of the type IIA sugra equations of motion,
although it is not supersymmetric at all. This is another example
of supersymmetry without supersymmetry, 
as discussed in section [\ref{susywsusy}]
Furthermore, this reduction did not produce the mentioned
configuration of D6 branes inside a Calabi-Yau. As can be
seen directly from the metric \bref{mal}, the six-dimensional
space spanned by the four-cycle and $(\theta,\phi)$ has the
structure of a  direct product $\C{K}_4 \times S^2$ instead
of being a fibration over the four-cycle. It cannot therefore
be the claimed Calabi-Yau three-fold.

To avoid such phenomenon, one can try to reduce along other $U$(1) isometries,
like the one generated by the Killing vector $\partial_{\phi}$. In order
to do so, the metric \bref{kahler1} must be rewritten in a form that makes
explicit the new $U$(1) fibration, \ie
\be\label{twist11d}
ds^2_{(11)}=dx_{0,2}^2+{3\over
2}(r^2+l^2)ds^2_{cycle}+U^{-1}dr^2+ {r^2 \over 4}\left( d\t^2 + m
B_{[1]}^2\right)
+\tilde{H}^{-1}(r,\theta) \left[d\phi-Uf^{-1} cos\t B_{[1]}\right]^2,
\ee
with the definitions
\be \label{defis}
 f(\t,r)\equiv sin^2\t+U(r) cos^2\t, \espai\espai
m(\t,r)\equiv\left(U^{-1}+cotg^2\t \right)^{-1},\ee\be
B_{[1]}\equiv d\psi+cos\t_1 d\phi_1+cos\t_2 d\phi_2, \espai\espai
\tH(r,\t)\equiv{4\over f(r,\t) r^2}.
\ee
Now both the metric and the supersymmetry parameters verify the
ansatz required to reduce down to IIA along $\phi$, and it yields
\be \label{goodiia}
ds^2_{IIA}=e^{2\Phi\over 3}\left(
dx_{0,2}^2+{3\over 2}(r^2+l^2)ds^2_{cycle}+U^{-1}dr^2+ {r^2
\over 4}  [ d\t^2 + m B_{[1]}^2]\right)
\ee
\be e^{4\Phi\over
3}={r^2\over 4} f(r,\t) \espai\espai C_{[1]}=-{Uf^{-1}
cos\t} B_{[1]}.
\ee
Note that all these background fields depend now on both
the radial coordinate $r$ and the angle $\theta$.
It can be checked that this configuration is a solution of the
equations of motion and that it has the expected four supersymmetries.

\subsection{Noncommutative Wrapped D6}

Just as we did in section 2, now we will look for a NC deformation
of the wrapped D6-branes, \bref{twist11d},
 by turning on a $B$-field, in this case,
along the $(x_1,x_2)$ plane.
As before, we explicitely break the worldvolume $SO(1,2)$ symmetry to
$R\times SO(2)$. As in the unwrapped case, we will also make use of 
the fact that, in 11d, the factors in front of the 10d part of 
the metric and in front of the $U$(1) fiber are related 
through the lifting ansatz \bref{lift1}.
Therefore, our ansatz for the bosonic fields is \footnote{We use the
definitions of \bref{defis} for the functions $f(r,\t)$ and $m(r,\t)$.}
\be ds^2_{(11)}=\tau^2(r,\t)\left[-dx_0^2+\sigma^2(r,\t)
dx_{1,2}^2+{3\over 2}(r^2+l^2)ds^2_{cycle}+U^{-1}dr^2+ {r^2
\over 4}\left( d\t^2 + m B_{[1]}^2\right)\right] +\ee\be
+\tau^{-4}(r,\t) \,\tH^{-1} \,\left[d\phi-Uf^{-1} cos\t
B_{[1]}\right]^2 \ee

\be
A_{[3]}=\chi(r,\t) \,\, dx^1\wedge dx^2\wedge \left[d\phi-Uf^{-1} cos\t B_{[1]}\right].
\ee
Note that we allow the functions of the ansatz to depend on $\t$.
Now we proceed to make an ansatz for the spinor.
Just like in the NC flat case, we expect that we will obtain a projection
signaling a bound state of MKK-M5
\be \label{const1}
\left(cos\a \G_{D6} + sin\a \tilde{\G}_{D4} \right) \e = \e,
\ee
for some angle $\a(r,\t)$ to be determined. Notice that since now the $B$-field
will be along $(x^1,x^2)$, we expect the D4 to span the directions 
$\{x^0x^3x^4x^5x^6\}$, so that $\tilde{\G}_{D4}=\G_{03456T}$.
As in the unwrapped case, see \bref{gamma1} \bref{analog}, equation \bref{const1}
implies 

\be
\e(r,\t,\phi,\psi)=
e^{{\a(r,\t)\over 2}\G_{12T}} \te(r,\t,\phi,\psi) \espai\espai
with \espai\espai \G_{D6}\,\te\,=\,\te.
\ee
Now we are ready to obtain the BPS equations by imposing that the
supersymmetry variation of the gravitino vanishes \bref{gravitino}.
The most immediate relations come from making them compatible for $A=0$ and $A=1,2$
and give
\footnote{We use primes for $\partial_r$ and
dots for $\partial_{\t}$. Also, the integration constant is set to one
in order to recover the commutative case when the three-form vanishes.}
\be  3\,{\tau' \over \tau}+{\sigma'\over \sigma}=0,
\espai \espai\espai  3\,{\dtau \over \tau}+{\dsigma\over \sigma}=0,
\ee
whose integration yields $\sigma=\tau^{-3}$.
The $A=5,6$ equations imply
that
\be \label{firsts}
\left(\partial_{\psi}+{\G_{36}\over 2}\right)\e=0 \espai\espai and \espai\espai
\G_{36}\,\e=\G_{45}\,\e.
\ee
while the $A=7$ equation implies that
\be
\tau^{-6}+\tH\tau^6\chi^2\,=\,1 \,.
\ee
Now taking a linear combination
of the $A=1,3,9$ equations, and assuming that the spinor
does not depend on the fiber coordinate $\phi$, 
one reaches another constraint similar to
\bref{gamma1}
\be \label{const2}
\left( cos\b \G_{3689}+sin\b \G_{3679} \right)\e=-\e,
\ee
with
\be
cos\b= U^{\undos}f^{-\undos}cos\t, \espai\espai\espai
sin\b=f^{-\undos}sin\t.
\ee
Since the matrices $\G_{3689}$ and $\G_{3679}$ anticommute, we
can proceed as in \bref{analog}, and rewrite 
this equation as 
\be
e^{-\beta \G_{78}}\e=-\G_{3689}\e,
\ee
whose most general solution is
\be
\e(r,\t,\psi)=e^{{\a(r,\t)\over 2}\G_{12T}} e^{{\b(r,\t)\over 2}\G_{78}}
\tte(r,\t,\psi) \espai\espai
with \espai\espai \G_{D6}\,\tte\,= -\G_{3689}\,\tte\,=\,\tte.
\ee
Plugging this into \bref{firsts} allows us to write
down the final ansatz for the spinor:
\be
\tte(r,\t,\psi)=\g(r,\t)
e^{-{\psi\over 2}\G_{89}}\e_0 \espai\espai with \espai\espai
\G_{D6}\e_0= -\G_{3689}\,\e_0\,=\,\e_0.
\ee

The first order BPS equations are
\be
6 {\tau'\over \tau}+{\chi'\over \chi}+{\tH ' \over \tH}\,=\,0
\ee\be
\dot{\a}-\undos \tH^{\undos}\tau^6 \dot{\chi}=0 \espai\espai
\a'-\undos \tH^{\undos}\tau^6 \chi'=0 \ee\be
{\g'\over \g} -{\tau' \over 2\tau}=0 \espai\espai
{\dot{\g}\over \g} -{\dot{\tau} \over 2\tau}=0
\ee
They can be solved analytically and, after fixing the integration
constants to reproduce the commutative case when $A_{[3]}$ vanishes, one
obtains
\be
\tau=\th^{1\over 6} \zespai\zespai \chi=-{\T \over \tH \th} \zespai\zespai
\g=\th^{1\over 12} \zespai\zespai  cos\a=-\th^{-\undos}\zespai\zespai
sin\a=-\T (\tH \th)^{-\undos},
\ee
with 
\be
\th(r,\t)=1+\Theta^2 \tH^{-1}(r,\t).
\ee
So the whole solution for the metric, three-form and Killing spinor is
\be
\label{ncd6-11}
ds^2_{(11)}=\th^{1\over 3} \left(-dx_0^2+ \th^{-1}
dx_{1,2}^2+{3\over 2}(r^2+l^2)ds^2_{cycle}+U^{-1}dr^2+
{r^2 \over 4} [ d\t^2 + m B_{[1]}^2]\right)
+ \ee\be
+\th^{-{2\over 3}} \tH ^{-1} \left(d\phi
 -{U f^{-1}cos\t} B_{[1]} \right)^2
\ee\be
A_{[3]}=-{\T\over \tH \th} \, dx^1\wedge dx^2 \wedge\left(
d\phi
 -{U f^{-1}cos\t} B_{[1]} \right)
\zespai\zespai
\e(r,\t,\psi)=\th^{1\over 12}(r,\t)e^{{\a(r,\t)\over 2}\G_{12T}} e^{{\b(r,\t)\over 2}\G_{78}}
e^{-{\psi\over 2}\G_{89}}\e_0  \label{bigspin}
\ee
with the constant spinor $\e_0$ subject to the following independent constraints
\be
\G_{D6}\,\e_0\,= \,\e_0, \espai\espai
\G_{36} \,\e_0 \,= \,\G_{45} \,\e_0 \, = \,\G_{89} \,\e_0.
\ee
Note that the introduction of the $B$-field has not
broken any extra supersymmetry, and the configuration
still preserves 4 real supercharges. The metric \bref{ncd6-11} and the 
three-form \bref{bigspin}  should be the supergravity
dual of a 2+1 \CN=2 $U$(N) field theory with noncommutativity along
the $(x^1,x^2)$ plane. As in \cite{Gomis:2001vg}, the field content
should consist of a vector multiplet.

Note also that this solution provides an example of M-theory
compactification with fluxes. The topology is $\MR^3\times \mathbb{X}_8$,
with $\mathbb{X}_8$ the non Ricci-flat internal manifold.
$\mathbb{X}_8$ consists of a complicated four dimensional fibration over
the \ka base space $S^2\times S^2$.
Remarkably, we can smoothly send to zero the noncommutativity, so that
the $A_{[3]}$ flux goes to zero and $\mathbb{X}_8$ becomes an $SU$(4)-holonomy Calabi-Yau
four-fold. From a $IIA$ perspective it describes a non-threshold bound
state of D6-D4 branes with the $D4$  wrapped around
the four-cycle, so that the arrays are

\begin{center}
\be
\begin{array}{c | c c c c c c c c c c}
IIA &x^0 & x^1 & x^2 & \t_1 & \t_2 & \phi_2 & \phi_1 & r &\t &
\psi \\ \hline
D6& -&-&-&-&-&-&-&&& \\
D4&-& & &-&-&-&-&&&\\
\end{array}\ee
\end{center}

\begin{center}
\be\begin{array}{c | c c c c c c c c c c c c} 
11d &x^0 & x^1 & x^2 & \t_1 & \t_2 & \phi_2 & \phi_1 & r &\t & \psi & \phi
\\ \hline
MKK& -&-&-&-&-&-&-&&& &\\
M5&-& & &-&-&-&-& & & & -\\
\end{array}\ee
\end{center}

\avall

\subsection{Compactifications to type IIA and to 8d gauged sugra}

In this subsection we will apply to the NC wrapped configuration
\bref{ncd6-11} the same arguments of  section [\ref{susywsusy}]
in order to discuss which compactifications preserve supersymmetry.

Since the susy parameters
\bref{bigspin} depend only $(\t,\psi)$, a $U$(1) reduction along $\phi$
will produce a type IIA solution preserving the four supercharges.
Explicitely
\be \label{nc-metrica}
ds^2_{IIA}=e^{2 \Phi/3}\th^{1\over 3} \left(-dx_0^2+ \th^{-1}
dx_{1,2}^2+{3\over 2}(r^2+l^2)ds^2_{cycle}+U^{-1}dr^2+
{r^2 \over 4} [ d\t^2 + m B_{[1]}^2]\right)
\ee\be
e^{4\Phi/3}=\th^{-{2\over 3}} \tH ^{-1} \espai\espai
B_{[2]}=-{\T\over \tH \th}\, dx^1\wedge dx^2 \ee\be
C_{[1]}=-Uf^{-1}cos\t\,B_{[1]} \espai\espai
C_{[3]}=-{\T\over \tH \th}Uf^{-1}cos\t\,\,\, dx^1\wedge dx^2 \wedge B_{[1]}.\ee
Instead, if we reduced along the $\partial_{\psi}$ isometry, we would
break all the supersymmetry.

On the other hand, when reducing to 8d gauged sugra, we find
a big difference between the commutative \bref{kahler1}
and the NC \bref{ncd6-11} cases, so we analyse them separately. 
As discussed in section [\ref{susywsusy2}], to see if supersymmetry
will be preserved in the $SU$(2) compactification, we 
have to transform the spinors to the $SU$(2) left-invariant 
$w$-base \bref{vi3}. To do so, we need to apply 
the rotation \bref{rotation} to the Killing spinors.

If we are in the commutative case, it is easy to see that the corresponding
spinors \bref{mmm} become constant, independent of all the $S^3$ angles. 
Therefore, the compactification can be performed preserving all four supersymmetries.
This is what allowed the authors of \cite{Gomis:2001vg} to find such solution
using 8d supergravity.

On the other hand, in the NC case, it can be checked that not even the
metric can be put in a form that satisfies the reduction ansatz,
so the compactification is simply not possible.
As a consequence, the NC wrapped D6 solution \bref{ncd6-11} could have
never been found with the usual gauged supergravity method.

\section{Conclusions}

We have constructed a supersymmetric configuration of
D6-branes wrapping a \ka four-cycle,
with a non-vanishing background $B$-field \bref{nc-metrica}.
We have shown that it can be interpreted as a
non-threshold bound state of D6-D4 branes, with the D4 dissolved in the
worldvolume of the D6, and  wrapped around the cycle.
The problem has been analysed in 11d because, as we have shown, its compactification
to 8d gauged sugra would have destroyed all the supersymmetries.
Similarly, the reduction to type IIA along a $U$(1) isometry
required some care to avoid this phenomenon.
We note that the correct reduction produces a background where
the fields depend on two transverse coordinates and it has
therefore cohomogeneity two.

Despite the fact that it preserves four supercharges,
the resulting metric is not Ricci-flat, so it does not have
reduced holonomy. This is the usual situation for compactifications with
background fluxes and, indeed, it would be nice to show that the four-cycle is
calibrated in the sense of generalised calibrations \cite{Gutowski}.

This configuration is expected to be dual in the IR of an \CN=2 NC gauge
theory in 2+1 dimensions, with noncommutativity along the spatial directions.
Such field theory is nonlocal in space, but local in time and it does not
suffer from unitarity or causality problems. 
This solution 
will hopefully allow the study of non-perturbative properties of this 
NC field theory.

\vskip 6mm

{\it{\bf Acknowledgements}}

We would like to thank Toine Van Proeyen and Daniela Zanon for collaboration
at earlier stages of this paper.
We are also grateful to 
Roberto Casalbuoni, Jerome Gauntlett, 
Jaume Gomis, Giorgo Longhi, David Mateos, Carlos N\'u\~nez, 
Tomas Ort\'{\i}n, Alfonso Ramallo, Joan Sim\'on 
and Paul Townsend for useful discussions. J.B. is supported
by a grant from Ministerio de Ciencia y Tecnolog\'{\i}a.
This work is partially supported by MCYT FPA, 2001-3598,
 and CIRIT, GC 2001SGR-00065.
T.M. is supported by a grant from the Commissionat per
a la Recerca de la Generalitat de Catalunya.

\end{document}